\documentclass[a4paper]{llncs}

\usepackage{graphicx}
\usepackage{times,verbatim}
\usepackage[UKenglish]{babel}
\usepackage{url}

\urldef{\mailsa}\path|amparo@iec.csic.es|
\urldef{\mailsb}\path|pcaballe@ull.es|

\begin{document}
\pagestyle{empty}

\mainmatter

\title{Weak Equivalents for Nonlinear Filtering Functions}

\titlerunning{Weak Equivalents for Nonlinear Filtering Functions}
\author{A. F\'uster-Sabater\inst{1} \and  P. Caballero-Gil\inst{2}}
\institute{Information Security Institute (CSIC), Serrano 144, 28006 Madrid, Spain \\
\email{amparo@iec.csic.es} \\
\and University of La Laguna, 38271 La Laguna, Tenerife,
Spain\\
 \email{pcaballe@ull.es}}
\maketitle

\begin{abstract}
In this paper we investigate equivalence of nonlinear filter functions applied to a Linear Feedback Shift Register (LFSR).
It is known that given a binary sequence generated from a nonlinear filter function applied to an LFSR, the same sequence can be generated from any other LFSR of the same length by using another filter function. However, no solution has been found for the problem regarding the issue of the computation of such an equivalent.
This paper analyses the specific case in which a reciprocal LFSR is used to generate an equivalent of the original nonlinear filter. 
The main advantage of the contribution is that weaker equivalents can be computed for any nonlinear filter, what could be used to cryptanalyze apparently secure generators.
Consequently, to evaluate the cryptographic resistance of a generator, the weakest equivalent cipher should be determined and not only a particular instance.
\end{abstract}


\section{Introduction}

A binary additive stream cipher is a synchronous cipher in which the binary output of a keystream generator is added bitwise to the binary plaintext sequence producing the binary ciphertext. The main goal in stream cipher design is to produce random-looking sequences that are unpredictable in an efficient way. From a cryptanalysis point of view, a good stream cipher should be resistant against known-plaintext attacks.

Most known keystream generators are based on Linear Feedback Shift
Registers (LFSRs)
\cite{Golomb}, whose output sequence is the image of a linear function applied to its successive states. Under certain conditions, this structure produces sequences with highly desirable features for cryptographic application. In particular, if its characteristic  polynomial is primitive, then the generated sequence, the so-called \emph{m}-sequence, exhibits certain useful properties such as a large period, good statistical distribution of 0's and 1's or excellent autocorrelation. However, the sequence produced by a LFSR  must never be used as keystream sequence in a stream cipher as the inherent linearity of this structure could be easily used to break the cipher.

An interesting  LFSR-based keystream generator is the nonlinear filter generator, which produces the keystream sequence as the image of a nonlinear Boolean function applied to the states of an LFSR. In particular, the nonlinear filter generator here analyzed (see Fig. 1) consists of two parts. 
\begin{enumerate}
  \item A LFSR with length $L$, characteristic polynomial $P(x) =  x^L + c_1\cdot x^{L-1} + \cdot \cdot \cdot  +  c_{L-1} \cdot x + c_L  $ with binary coefficients that, from an initial state $IS$, generates an output sequence $\{a_n \}$ .
  \item A nonlinear Boolean function $F: GF(2)^L \rightarrow GF(2)$, called filter function, whose inputs variables are the successive \emph{L}-bit states of the LFSR and whose image is the binary keystream sequence $\{z_n \}$.
\end{enumerate}
Although the sequences produced by LFSRs have been well studied, the same does not apply to the sequences obtained with nonlinear filter generators.

\begin{figure}[!t]
	\centering
		\includegraphics[width=2.5in]{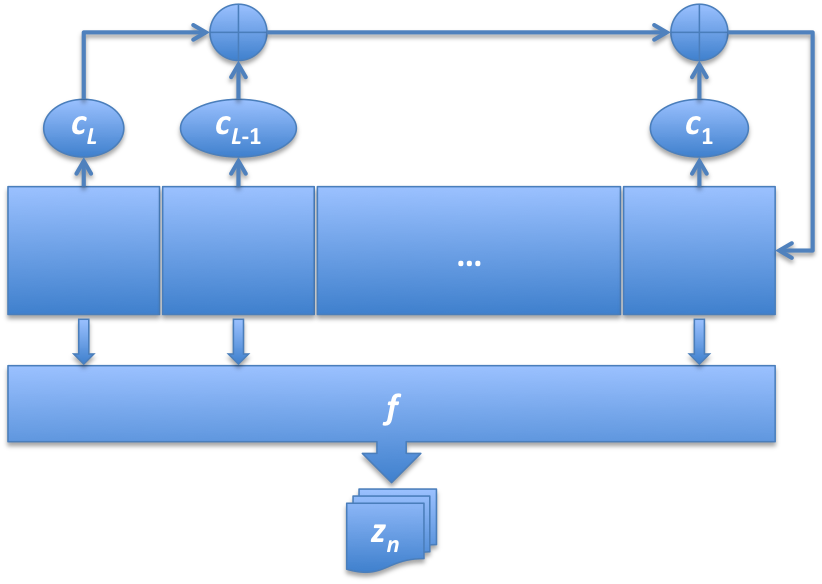}
	\caption{A Nonlinear Filter Generator}
\end{figure}

This work deals with the relationship between different nonlinear filter generators that produce exactly the same sequence. The main goal is to show that although the study of the generator's properties leads to right conclusions about the properties of the generated sequence, sometimes misleading inferences can be drawn. In particular, this paper shows that two structures with apparently different security levels can produce the same keystream sequence. Indeed, this result can be seen as a proof that the actual security level of a generator is the security level of the weakest element in its corresponding class.

This paper is organized as follows. Section 2 includes a succinct revision of related works. In Section 3 after some necessary preliminaries, the general problem of counting equivalent nonlinear filter generators is addressed as well as the relationship among them is studied. Afterwards, Section 4 provides a brief explanation of the proposal based on the new concept of reciprocal filter generators and introduces a novel method for computing weaker equivalent nonlinear filter generators through a pedagogical example. Finally, Section 5 discusses conclusions and possible future research lines.

\section{Related Work}

A useful tool to study  binary sequences is the Berlekamp-Massey algorithm \cite{BM}, which determines the shortest LFSR that generates any input finite binary sequence. The length of such a LFSR is the linear complexity of the sequence. General lower and upper bounds on the linear complexity of filtered sequences have been derived in the works \cite{Key76} and \cite{Fuster95}, while better lower bounds can be found in \cite{Rue86} and \cite{Sch99} for special cases.

The eSTREAM project \cite{estream} is the most significant effort for designing secure stream ciphers. It was a multi-year project whose objective was to promote the design of efficient stream ciphers suitable for widespread adoption. As a result of the project, a portfolio of seven stream ciphers with two different profiles, software and hardware, was published. One of them, the so-called SOSEMANUK, is an LFSR-based generator where the length of the used LFSR is 10 and the content of each stage is an element of $GF(2^{32})$. Such a generator uses design principles similar to the stream cipher SNOW 2.0 that led to the later called SNOW 3G, which  forms the heart of the 3GPP confidentiality and integrity algorithms  for LTE and LTE-Advanced \cite{Snow3G}.

Several references of cryptanalytic attacks on nonlinear filters can be found.

The basic correlation attack against the nonlinear filter generator was published in \cite{Sie85}, where correlations between the  filtered sequence $\{z_n \}$ and the LFSR \emph{m}-sequence $\{a_n \}$ are used to build an equivalent generator consisting of a nonlinear combination of several LFSRs. The main drawbacks of this attack is the huge amount of time needed for computing the necessary  correlations and the requirement that the filter function $F$ must have high correlation to an affine function. After defining the nonlinearity of a Boolean function as the minimum Hamming distance between such a function and an affine function, a practical consequence of the basic correlation attack is that the cryptographic designer has to choose  highly nonlinear filter functions for  the nonlinear filter generators. Afterwards, the concept of basic correlation attacks was improved by the  fast correlation attack  described in \cite{MS89}. Two common disadvantages of the different versions of these attacks are: a) the large number of intercepted keystream bits needed to perform a successful attack and b) the assumption that the filter function is not highly nonlinear.

A general inversion attack was proposed in \cite{GCD99} for any filter function. Consequently, an easy characterization was obtained for  filter generators that are  resistant against the inversion attack. On the other hand, the works \cite{GR94} and \cite{Fil00} proposed the so-called decimation attack for any LFSR-based keystream generator. The idea is to consider a decimated sequence of the intercepted keystream sequence so that the decimated sequence can be generated from a decimated LFSR sequence. However, according to \cite{Rue86} if the LFSR's length  $L$ is a prime number, then  the decimation attack provides no further advantages.

In the last years, several algebraic attacks on stream ciphers have been published. In these attacks, the attacker uses the bits of the intercepted sequence to set a nonlinear system of polynomial equations in terms of the LFSR bits. The main issue regarding these types of attacks is that, as shown in \cite{GJ79}, the problem of solving a nonlinear system of multivariate equations is NP-hard even if all the equations are quadratic and the underlying field is $GF(2)$. In order to deal with this, the so-called  algebraic method XL \cite{CKPS00} was proposed  to solve the nonlinear system of quadratic equations for certain nonlinear filter generators. In order to make them resistant against this attack, the filter function should be not only highly nonlinear, but also have a large distance to approximations of low algebraic degree.

The so-called time-memory-data tradeoff attacks \cite{BS00} can be easily prevented in nonlinear filter generators by using LFSRs with large length. There is another interesting attack, the so-called guess and determine attack \cite{G&D}, which exploits the relationship between internal values (such as the recurrence relationship in the LFSR), and the relationship used to construct the keystream sequence from the internal values. As its name indicates, this attack proceeds by guessing some internal values and then using the relationships to determine other internal values. After such an attack,  the encryption is considered broken when a complete internal state has been determined from the guessed values. This type of attack can be prevented by choosing adequate polynomials of the LFSRs.

One of the most closely related papers to this one is \cite{Loh01}, where the so-called  linear transformation attack against the nonlinear filter generator was proposed.
The idea behind this attack is to transform the given generator into an equivalent nonlinear filter generator with the same LFSR but a filter function that is  better suited for some of the above described attacks. 

Another close paper is \cite{RonjomCid}, where the authors define an equivalence class of nonlinear filter generators, showing that a number of important cryptographic properties are not invariant among elements of the same equivalence class. The authors themselves acknowledge that determining the weakest equivalent cipher is a very difficult task because the size of the equivalence class is very large. In this paper we do not deal with the complete nonlinear equivalence class, but only with  one of its members, the one we have identified that in many cases leads to a weaker equivalent generator.

In conclusion, each  attack against nonlinear filter generators leads to one or more than one conclusion about desirable properties of the LFSR and/or of the filter function. Consequently, one of the main research issues regarding nonlinear filter generators is how to construct a good Boolean function to achieve resistance against all the aforementioned attacks. This work deals with this issue because it proves that the properties of the generator not always guarantee security at the output sequences.

\section{General Study of Equivalent Filters}

In this section, first of all the number of equivalent filters is obtained then the relationship between them is analyzed.

For a filter generator consisting of an LFSR with characteristic polynomial $P_1(x)$ and a filter function $F_1(x)$, it is always possible to generate the same sequence with any other LFSR of the same length and another filter function.

 Let $\alpha$ be a root of the characteristic polynomial $P_1( x )$ as well as a primitive element of $GF (2^L )$. In that case, if $gcd ( k, 2^L- 1) = 1 $, then $\alpha ^k$ is also a primitive element of $GF ( 2^L )$, so there are $\phi ( 2^L- 1) $ primitive elements of $GF ( 2^L )$. In particular, the $L$ conjugates of any element (which are the successive square powers), e.g. $\alpha, \alpha ^2, \alpha ^4, \alpha ^8,..., \alpha ^{2^{L-1}}$, are primitive elements of $GF ( 2^L )$ as well as roots of the same polynomial, which can be computed by the expression 
$\prod_ { i = 0 } ^ {L -1}  ( x - \alpha^{2^i})$ in $ GF (2^L )$.

Therefore, there are $\phi ( 2^L- 1)/L $ primitive polynomials of $GF ( 2^L ) $, each one with $L$ roots that are all conjugates of a primitive element. Since each one of these polynomials defines an LFSR of length $L$, there are $\phi ( 2^L- 1)/L $ different LFSRs of length $L$, each of them corresponding to a set of conjugates of a primitive element of $GF ( 2^L )$.

In conclusion, since any binary sequence obtained with a nonlinear filter can be generated by a filter function over each LFSR, then there are $\phi ( 2^L- 1)/L $  different nonlinear filter generators that can be used to generate it.

The relationship between two primitive elements, $\alpha$ and $\beta$, roots of two different characteristic polynomials of two different LFSRs of length $L$ is given by the expression $ \beta = \alpha ^k$ being $gcd ( k, 2^L- 1) = 1$ and $ k \neq 2^i \cdot j $ (mod $2^L- 1)$ with $i, j > 0$.

This knowledge on  the relationship between the characteristic polynomials $P_1 ( x )$ and $P_2 ( x )$ of two LFSRs could help to define the relationship between two filter functions $F_1 ( x )$ and $ F_2 ( x )$, which  are part of two equivalent generators that produce the same filtered sequence.

  \begin{table}[!ht]
  \begin{center}
  \begin{tabular}{|c||c|c|c|c|}
   \hline
  $L$&3&4&5&6 \\ \hline
$2^L -1$& 7& 15& 31& 63  \\ \hline
N. filters&2&2&6&6 \\ \hline
$k$ defining filters&1,2,4& 1,2,4,8 &1,2,4,8,16&1,2,4,8,16,32 \\
&3,5,6&7,11,13,14&3,6,12,24,17&5,10,20,40,17,34 \\
&&&5,10,20,9,18&11,22,44,25,50,37\\
&&&7,14,28,25,19&13,26,52,41,19,38\\
&&&11,22,13,26,21&23,46,29,58,53,43\\
&&&15,30,29,27,23&31,62,61,59,55,47\\
  \hline
 \end{tabular}
 \end{center}
 \caption{Examples of counting of equivalent filters}
  \end{table}

As we can see in Table 1, the cases $ k = 1$ and $ k = 2^{L -1}-1 $ always determine different sets of conjugate roots that define different LFSRs. In fact, the corresponding polynomials for the roots $\alpha$ and $\beta = \alpha ^ {2 ^ { L - 1} -1 }$ are always reciprocal.

Any \emph{m}-sequence $\{ a_n \} $ can be written in terms of the roots of the characteristic polynomial of the LFSR through the trace function, so that $a_n = Tr ( \alpha ^ n) = \sum_ { i = 0 } ^{ L -1} \alpha ^{ n2 ^ i }$.  Consequently, given a sequence $\{ a_n \} $ generated by a LFSR with polynomial
$P_1 ( x )$ and root $\alpha$ and another sequence $\{ b_n \} $ generated by other LFSR with polynomial $P_2 ( x )$ and root $\beta$ such that $\beta = \alpha ^k$, we have that $a_n = \sum_ { i = 0 } ^{ L -1} \alpha ^{ n2 ^ i }$ and $b_n = \sum_ { i = 0 } ^{ L -1} \alpha ^{ kn2 ^ i }$. This is shown with an example in Table 2.

  \begin{table}[!ht]
  \begin{center}
  \begin{tabular}{|c||c|c|c|c|}
   \hline
Roots&Polynomial&m-sequence \\  \hline \hline
$\alpha, \alpha ^2, \alpha ^4, \alpha ^8, \alpha ^{16}$&$x^5 + x^4 + x^3 + x^2 + 1$& $\{ a_n \} $\\  \hline
$\alpha ^{15}, \alpha ^{30}, \alpha ^{29}, \alpha ^{27}, \alpha ^{23} $ &reciprocal=&$\{ b_n \} $ reverse of $\{ a_n \} $\\
&$=x^5 + x^3 + x^2 + x + 1$&\\\hline
$\alpha ^3, \alpha ^{6}, \alpha ^{12}, \alpha ^{24}, \alpha ^{17} $ &$\prod_{i = 0}^4 (x- \alpha ^{3 \cdot 2^i})=$& $\{ c_n \} $\\
&$=x^5 + x^4 + x^2 + x + 1$& \\ \hline
($\alpha ^3)^{15}= \alpha ^{14}, \alpha ^{28}, \alpha ^{25}, \alpha ^{19}, \alpha ^{7}$&reciprocal=& $\{ d_n \} $ reverse of $\{c_n \} $\\
&$x^5 + x^4 + x^3 + x + 1$&\\\hline
$\alpha ^5, \alpha ^{10}, \alpha ^{20}, \alpha ^{9}, \alpha ^{8} $& $\prod_{i = 0}^4 (x- \alpha ^{5 \cdot 2^i})=$&$\{ e_n \} $\\

&$x^5 + x^3 + 1$&\\ \hline

 ($\alpha ^5)^{15}= \alpha ^{13},\alpha ^{26}, \alpha ^{21}, \alpha ^{11}, \alpha ^{22}$&reciprocal= &$\{ f_n \} $ reverse of $\{e_n \} $\\
&$x^5 + x^2 + 1$&\\\hline

 \end{tabular}
 \end{center}
 \caption{Examples of relationships between roots, polynomials and \emph{m}-sequences}
  \end{table}

If two filter generators defined by the corresponding polynomials and  filter functions $( P_1 ( x ), F_1 ( x ) )$ and $( P_2 ( x ), F_2 ( x ) )$ generate the same sequence, then we have that:

$$ F_1 (a  _n, a_ {n +1}, ..., a_{n + L -1} ) = F_1 ( \sum_ {i = 0} ^{ L -1} \alpha ^{ n2^ i },  \sum_ {i = 0} ^{ L -1} \alpha ^{ (n+1)2^ i },...,  \sum_ {i = 0} ^{ L -1} \alpha ^{( n+L-1)2^ i })=$$
$$ F_2 (b  _n, b_ {n +1}, ..., b_{n + L -1} ) = F_2 ( \sum_ {i = 0} ^{ L -1} \alpha ^{ kn2^ i },  \sum_ {i = 0} ^{ L -1} \alpha ^{ k(n+1)2^ i },...,  \sum_ {i = 0} ^{ L -1} \alpha ^{k( n+L-1)2^ i }).$$

The Algebraic Normal Form of Boolean functions allows us to write the sequence generated by a filter generator $(P_1 ( x ), F_1 ( x ) )$  in terms of a root $\alpha$ of the polynomial $ P_1 ( x )$ and binary coefficients, as follows:

$$ F_1 (a  _n, a_ {n +1}, ..., a_{n + L -1} ) =$$$$=c_0 a_n + \cdot \cdot \cdot + c_{L - 1} a_{ n + L -1} + c_{0, 1} a_n a_{n +1}+ \cdot \cdot \cdot  + c_{L - 2, L -1} a_{n + L-2}  a_{n + L -1} +$$$$+\cdot \cdot \cdot + c_{0, 1, ... L -1} a_n a_{n + 1} \cdot \cdot \cdot a_{n + L -1} =$$
$$ =c_0  \sum_ {i = 0} ^{ L -1} \alpha ^{ n2^ i } + \cdot \cdot \cdot + c_{L - 1}  \sum_ {i = 0} ^{ L -1} \alpha ^{ (n+L-1)2^ i } + c_{0, 1} \sum_ {i = 0} ^{ L -1} \alpha ^{ n2^ i }  \sum_ {i = 0} ^{ L -1} \alpha ^{ (n+1)2^ i }+ \cdot \cdot \cdot  +$$$$+ c_{L - 2, L -1}  \sum_ {i = 0} ^{ L -1} \alpha ^{ (n+L-2)2^ i }   \sum_ {i = 0} ^{ L -1} \alpha ^{ (n+L-1)2^ i } +$$$$+\cdot \cdot \cdot + c_{0, 1, ... L -1}  \sum_ {i = 0} ^{ L -1} \alpha ^{ n2^ i }  \sum_ {i = 0} ^{ L -1} \alpha ^{ (n+1)2^ i } \cdot \cdot \cdot  \sum_ {i = 0} ^{ L -1} \alpha ^{ (n+L-1)2^ i }.$$

Thus,  if the expression is partitioned into cosets (sets of integers $E  \cdot 2^i$ $ mod ( 2^L- 1)$ with $0 \leq i \leq L -1)$, then the function can be expressed as:

$$F_1 (a  _n, a_ {n +1}, ..., a_{n + L -1} ) =
  \sum_ {i = 0} ^{ L -1} C_{coset1}\alpha ^{ n \cdot coset1 \cdot 2^ i } +
C_{coset2}\alpha ^{ n \cdot coset2 \cdot 2^ i } +
\cdot \cdot \cdot $$
  with $C_{cosetj} \in GF(2^L)$.

The weights of the cosets whose coefficients are nonzero in the previous expression provide some information about the function, i.e. its order.

In particular, if the relationship $\beta = \alpha ^k$ between two filter generators $ (P_1 ( x ), F_1 ( x ) )$ and $( P_2 ( x ), F_2 ( x ) )$ that generate the same sequence is known, then:

 $$
F_2 (b  _n, b_ {n +1}, ..., b_{n + L -1} ) =
  \sum_ {i = 0} ^{ L -1} D_{coset1}\alpha ^{ n \cdot coset1 \cdot 2^ i } +
D_{coset2}\alpha ^{ n \cdot coset2 \cdot 2^ i } +
\cdot \cdot \cdot $$
  with $D_{cosetj} \in GF(2^L)$.

Then, the cosets that appear in both expressions must be paired so that for each coset $cosetv$ in the first expression, another coset $cosetw$ exists in the second expression. That is:
 $$
\sum_ {i = 0} ^{ L -1} C_{cosetv} \alpha ^{ n \cdot cosetv \cdot 2^ i } =
\sum_ {i = 0} ^{ L -1} D_{cosetw}\alpha ^{ n \cdot cosetw \cdot 2^ i } .$$

\section{Reciprocal Filters}

From the results shown in the previous Section, if two LFSRs with reciprocal polynomials $P_1 ( x )$ and $P_2 ( x )$ are considered, two conclusions can be obtained.

 \begin{enumerate}
   \item If the same filter function $F ( x )$ is applied to both LFSRs, then different sequences are generated. The Berlekamp-Massey algorithm can be used on the resulting filtered sequences. In fact, from the factorizations of the obtained polynomials it can be concluded that they always correspond exactly to the same cosets.
   \item In order to generate the same sequence with those LFSRs, two different filter functions $F_1 ( x )$ and $F_2 ( x )$  must be used. Since the factorization of the polynomial obtained with the Berlekamp-Massey algorithm corresponds to mirrored complementary cosets in the groups defined by each of the LFSRs, the order of the filter functions are influenced by the weights of those cosets. In particular, it can be concluded  that $order (F_i) = max (L$-(weight of each coset linked to the factorization of the polynomial of the sequence)).
 \end{enumerate}

Thus, if there is a filter generator producing a sequence whose factorization only corresponds to cosets of weight $> L / 2$, then there is an equivalent filter that is less strong and has order  $< L / 2$. Regarding such an equivalent filter, it is a well known fact that the LFSR is the reciprocal of the original one.

If a filter function  has order $\sim L / 2$, since the order is given by the maximum of the weights of the cosets associated with the factorization, then there is an equivalent filter of order $\geq L / 2$ as such a degree is given by the maximum of the weights of the cosets. Consequently, if a reciprocal LFSR is used, then it is known that its weight is at least $L-L / 2$ . This can be seen as a proof of the known recommendation about using filter functions of order $\sim L / 2$.

From all the aforementioned, it can be concluded that for any filter generator, an equivalent filter generator called reciprocal filter can be always obtained to generate the same sequence.
In order to determine the reciprocal filter for any known filter generator, the proposed procedure includes the following four basic steps:
\begin{enumerate}
\item Determine the  relationships between the roots of the characteristic polynomials of the initial LFSR and its reciprocal.
\item Express both \emph{m}-sequences through the trace function.
\item Compute the  coefficients of the cosets in the expression of the filter function.
\item Solve a version of the discrete knapsack problem to build the reciprocal filter function.
\end{enumerate}

This procedure is illustrated through a pedagogical example.

{\bf Example:}

Given an LFSR of length $L=5$, characteristic polynomial $P_1(x) = x^5 + x^3 + 1$ and initial state  $IS_1 = (1,0,0,0,0) $, the filter function of order 4 $$F_1(a_0,a_1,a_2,a_3,a_4)=a_0a_1a_3a_4+a_0a_2a_3a_4+a_0a_1a_4+a_0a_1a_3+a_1a_3a_4+a_0a_3a_4+$$$$+   a_1a_2+a_1a_3+a_2a_4+a_0a_2+a_0a_3+a_1+a_2+a_3
$$
ia applied to produce the filtered sequence of period $2^5-1$,
$$00101 10110 10110 11100 00100 10101 1.$$

The reciprocal LFSR has characteristic polynomial $P_2(x) = x^5 + x^2 + 1$, whose root $\beta$  is related  to the root $\alpha $ of $ P_1 ( x )$ by the expression  $\beta =\alpha ^{2^{5-1}-1}=\alpha ^{15}$. Furthermore, thanks to the modular inverse of 15 (mod 31), the inverse relationship can be obtained $\alpha=\beta ^{29}.$

At the same time, the \emph{m}-sequences $\{a_n\}$ and $\{b_n\}$ obtained from $P_1(x)$ and $P_2(x)$, respectively, can be expressed by means of their trace expressions:
$$a_n=\alpha^n+\alpha^{2n}+\alpha^{4n}+\alpha^{8n}+\alpha^{16n}$$
$$b_n=\beta^n+\beta^{2n}+\beta^{4n}+\beta^{8n}+\beta^{16n}.$$

Consequently, the filter functions $F_1$ and  $F_2$ can be expressed in terms of the  $\phi ( 2^5- 1)/5 = 6$  cosets $\{15,11,7,5,3,  1\}$.
In particular,  the coefficient $C_{15}$ corresponding to the coset of maximum order 4  can be obtained by the root presence test \cite{Rue86}, while the coefficients $C_7$ and $C_{11}$ corresponding to the cosets of order 3 can be computed by grouping terms:
$$C_{15}=\alpha^6,C_7=\alpha^{24},C_{11}=\alpha^4.
$$
From these values, it can be concluded that no more cosets of lower weight appear in the expression of the filter function $F_1$. Thus,
$$F_1=C_{15} \alpha^{15n}+C_{15}^2 \alpha^{30n}+C_{15}^4 \alpha^{29n}+C_{15}^8 \alpha^{27n}+C_{15}^{16} \alpha^{23n}+$$$$+C_7 \alpha^{7n}+C_7^2 \alpha^{14n}+C_7^4 \alpha^{28n}+C_7^8 \alpha^{25n}+C_7^{16} \alpha^{19n}+$$$$+C_{11} \alpha^{11n}+C_{11}^2 \alpha^{22n}+C_{11}^4 \alpha^{13n}+C_{11}^8 \alpha^{26n}+C_{11}^{16} \alpha^{21n}.$$

If $\alpha=\beta ^{29}$ is substituted in that expression, then the filter function $F_2$  generating the same sequence can be expressed as:

$$F_2=C_{15} \beta ^{29 \cdot 15n}+C_{15}^2 \beta ^{29 \cdot 30n}+C_{15}^4 \beta ^{29 \cdot 29n}+C_{15}^8 \beta ^{ 29 \cdot 27n}+C_{15}^{16} \beta ^{29 \cdot 23n}+$$
$$+C_7 \beta ^{29 \cdot 7n}+C_7^2 \beta ^{29 \cdot 14n}+C_7^4 \beta ^{29 \cdot 28n}+C_7^8 \beta ^{29 \cdot 25n}+C_7^{16} \beta ^{29 \cdot 19n}+$$$$+C_{11} \beta ^{29 \cdot 11n}+C_{11}^2 \beta ^{29 \cdot 22n}+C_{11}^4 \beta ^{29 \cdot 13n}+C_{11}^8 \beta ^{29 \cdot 26n}+C_{11}^{16} \beta ^{29 \cdot 21n}=$$
$$=
C_{15} \beta ^{n}+C_{15}^2 \beta ^{2n}+C_{15}^4 \beta ^{4n}+C_{15}^8 \beta ^{8n}+C_{15}^{16} \beta ^{16n}+$$
$$+C_7 \beta ^{17n}+C_7^2 \beta ^{3n}+C_7^4 \beta ^{6n}+C_7^8 \beta ^{12n}+C_7^{16} \beta ^{24n}+$$
$$+C_{11} \beta ^{9n}+C_{11}^2 \beta ^{18n}+C_{11}^4 \beta ^{5n}+C_{11}^8 \beta ^{10n}+C_{11}^{16} \beta ^{20n}.
$$

Consequently, it can be concluded that only the cosets 1, 3 and 5 appear in this expression. Furthermore, their  coefficients $D_1$, $D_3$ and $D_5$ are given by:
$$D_{1}= C_{15}=\alpha^6=\beta ^{29 \cdot 6}=\beta ^{19}$$
$$D_{3}= C_{7}^2=\alpha^{24 \cdot 2}=\alpha^{17}= \beta ^{29 \cdot 17}=\beta ^{28}$$
$$D_{5}= C_{11}^4=\alpha^{4 \cdot 4}=\alpha^{16}= \beta ^{29 \cdot 16}=\beta ^{30}.$$

Since the maximum weight of the cosets in that expression is 2, the nonlinear terms of order 2 in the expression of $F_2$ are analyzed. As before, for each nonlinear term of order 2 the coefficients $D_3$ and $D_5$ can be obtained by the root presence test \cite{Rue86}, while the coefficient $D_1$  corresponding to the coset of weight 1 can be computed by grouping terms, as it is shown in Table 3.

  \begin{table}[!ht]
  \begin{center}
  \begin{tabular}{|c||c|c|c|}
   \hline
&$D_3$&$D_5$&$D_1$ \\ \hline
$\;b_0b_1\;$& $\;\beta ^{19}\;$& $\;\beta ^{30}\;$& $\;\beta ^{16}\;$  \\
$b_0b_2$& $\beta ^{7}$& $\beta ^{29}$& $\beta $  \\
$b_0b_3$& $\beta $& $\beta ^{19}$& $\beta ^{17}$  \\
$b_0b_4$& $\beta ^{14}$& $\beta ^{27}$& $\beta ^{2}$  \\
$b_1b_2$& $\beta ^{22}$& $\beta ^{4}$& $\beta ^{17}$  \\
$b_1b_3$& $\beta ^{10}$& $\beta ^{3}$& $\beta ^{2}$  \\
$b_1b_4$& $\beta ^{4}$& $\beta ^{24}$& $\beta ^{18}$  \\
$b_2b_3$& $\beta ^{25}$& $\beta ^{9}$& $\beta ^{18}$  \\
$b_2b_4$& $\beta ^{13}$& $\beta ^{8}$& $\beta ^{3}$  \\
$b_3b_4$& $\beta ^{28}$& $\beta ^{14}$& $\beta ^{19}$  \\ \hline

 \end{tabular}
 \end{center}
 \caption{Coefficients of the cosets 3, 5 and 1 for all the possible terms of order 2}
  \end{table}

An interesting version of the discrete knapsack problem defined by the coefficients in Table 3 is then solved so that for each one of the two first columns corresponding to the cosets of maximum weight, those elements whose sum coincides with the corresponding  known coefficient are computed. In particular, the solution shows that the coefficients corresponding to the products $b_0b_2$, $b_1b_2$, $b_1b_3$, $b_1b_4$ and $b_3b_4$ give both values:
$$D_3=\beta ^{7}+\beta ^{22}+\beta ^{10}+\beta ^{4}+\beta ^{28}=\beta ^{28}$$
$$D_5=\beta ^{29}+\beta ^{4}+\beta ^{3}+\beta ^{24}+\beta ^{14}=\beta ^{30}.$$

This result applied on the last column produces that, in order to obtain the final sum $D_{1}=\beta ^{19}$, the linear elements $b_1+b_2+b_4$ have to be included in the filter function $F_2$:
$$
D_1=\beta+\beta ^{17}+\beta ^{2}+\beta ^{18}+\beta ^{19}+\beta +\beta ^{2}+\beta ^{4} =\beta ^{19}.
$$

 Thus, the final expression of the equivalent filter function is obtained:
$$
F_2
(b_0,b_1,b_2,b_3,b_4)=b_0b_2+b_1b_2+b_1b_3+b_1b_4+b_3b_4+b_1+b_2+b_4.
$$

This function applied on the reciprocal LFSR with  characteristic polynomial $P_2(x) = x^5 + x^2 + 1$ and initial state  $IS_2 = (1,0,0,1,0) $, produces the
same filtered sequence of the input filter generator
$$00101 10110 10110 11100 00100 10101 1.$$

Recall that $F_2$ is a function of order 2 with the same number of terms of order 2 and 1 than $F_1$ but without terms of order 3 neither 4. Thus, from a cryptographic point of view, the attacker will find an easier attack against $F_2$ than against $F_1$ although both filters generate exactly the same sequence.

Consequently, this example shows that the proposed  method can be applied on any known filter generator  in order to produce an equivalent filter, which in the case of reciprocal LFSR is of a lower order. This is a proof that some generators apparently secure can have weaker equivalents, and what is more important, that these equivalents can be computed.

\section{Conclusions and Future Works}
This work has addressed the problem of computing equivalent nonlinear filters that produce the same sequence as a known filter generator. In particular, it  analyzes the case in which a reciprocal LFSR is used to define an equivalent nonlinear filter. In fact, under such conditions there are specific relationships between the two filter functions that allow the definition of a specific method for computing the equivalent filter function. The study concludes that the equivalent generator can have a security level that is lower than the one of the original filter. Therefore, the proposed method allows  building equivalents that are weaker than the starting filters. In conclusion, this work shows that two structures with apparently different security levels according to their properties, can produce exactly the same keystream sequence, so  both generators must be considered as insecure as the weakest one.

Given the difficulty of the subject, there are still many open issues. In particular, one of them is the development of optimum methods for solving the particular knapsack problem that appears in the last phase of the proposed method. Also,  a  study similar to the one shown in this paper, but on other equivalents that do not correspond to the reciprocal LFSR, could be useful for potential cryptanalitic attacks on nonlinear filter functions.

\section*{Acknowledgment}

Research supported by the Spanish MINECO and the European FEDER Funds under projects TIN2011-25452 and IPT-2012-0585-370000.


\begin{thebibliography}{6}

\bibitem{BS00}  Biryukov, A.,   Shamir, A. Cryptanalytic time/memory/data tradeoffs
for stream ciphers.  Advances in Cryptology, ASIACRYPT’00, Lecture Notes in Computer Science 1976, pp. 1-–13.
Springer-Verlag, 2000.

\bibitem{CKPS00}  Courtois, N.,  Klimov, A.,  Patarin, J.,  Shamir, A.
Efficient algorithms for solving overdefined systems of multivariate polynomial equations. Advances in Cryptology, EUROCRYPT’00, Lecture Notes in Computer Science 1807, pp. 392–-407.
Springer-Verlag, 2000.

\bibitem{estream} eSTREAM: the ECRYPT Stream Cipher Project. Available from http://www.ecrypt.eu.org/stream/

\bibitem{Fil00}  Filiol, E. Decimation attack on stream ciphers. Advances in Cryptology, INDOCRYPT 2000, Lecture Notes in Computer Science 1977,
pp. 31–-42. Springer-Verlag, 2000.

\bibitem{Fuster95}  F´uster-Sabater A.,   Caballero-Gil, P. On the linear complexity
of nonlinearly filtered pn-sequences.  Advances in Cryptology, ASIACRYPT’94, Lecture Notes in Computer Science 917, pp.
80–-90. Springer-Verlag, 1995.

\bibitem{GR94} Games, R.A., Rushanan, J.J. Blind synchronization of m-
sequences with even span. Advances in Cryptology,
EUROCRYPT’93, Lecture Notes in Computer Science 765, pp. 168-–180.
Springer-Verlag, 1994.

\bibitem{GJ79}   Garey, M.R.,   Johnson, D.S. Computers and Interactability.
Freeman and Company, 1979.

\bibitem{GCD99}  Goli´c, J.D.,  Clark, A.,  Dawson, E. Generalized inversion attack on nonlinear filter generators. IEEE Transactions on Computers,
49(10), pp. 1100–-1109,  2000.

\bibitem{Golomb}Golomb, S.W., Shift Register-Sequences, Aegean Park Press, Laguna Hill, 1982.

\bibitem{Key76} Key, E.L. An analysis of the structure and complexity of nonlinear
binary sequence generators. IEEE Transactions on Information Theory,
22(6), pp. 732-–736,  1976.

\bibitem{Loh01}  L¨ohlein, B. Design and analysis of cryptographic secure keystream
generators for stream cipher encryption. PhD thesis, Faculty
of Electrical and Information Engineering, University of Hagen, Germany,
 2001.

\bibitem{BM} Massey, J. L.  Shift-register synthesis and BCH decoding, IEEE Transactions on Information Theory, IT-15 (1), pp. 122-–127, 1969.

\bibitem{MS89}  Meier W.,   Staffelbach, O.J. Fast correlation attacks on stream
ciphers. Journal of Cryptology, 1(3), pp. 159–-176, 1989.

\bibitem{G&D} Pasalic, E. On guess and determine cryptanalysis of LFSR-based stream ciphers.  IEEE Transactions on Information Theory, 55(7), pp. 3398--3406, 2009.

\bibitem{RonjomCid} Rønjom, S.,Cid, C. Nonlinear equivalence of stream ciphers.  Fast Software Encryption, pp. 40--54. Springer-Verlag, 2010.

\bibitem{Rue86} Rueppel, R.A. Analysis and Design of Stream Ciphers. Springer-Verlag,
1986.

\bibitem{Sch99}  Schneider, M. Methods of generating binary pseudo-random sequences
for stream cipher encryption. PhD thesis, Faculty of Electrical
Engineering, University of Hagen, Germany,  1999.

\bibitem{Sie85} Siegenthaler, T.  Decrypting a class of stream ciphers using ciphertext only. IEEE Transactions on Computers, 100(1), pp. 81--85, 1985.

\bibitem{Snow3G} SNOW 3G specification.  Specification of the 3GPP Confidentiality and Integrity Algorithms UEA2 and UIA2. Available from http://www.3gpp.org/DynaReport/35216.htm

\end{thebibliography}
\end{document}